\documentclass[aps,prd,twocolumn,groupedaddress,letterpaper,nofootinbib]{revtex4}

\usepackage{fullpage}
\usepackage{graphicx}
\usepackage{amsmath}
\usepackage{color}
\usepackage{comment}

\begin{document}

\title{Holographic bounds and finite inflation}

\author{Daniel Phillips}
\affiliation{Department of Physics, UC Davis}
\author{Andrew Scacco}
\affiliation{Department of Physics, UC Davis}
\author{Andreas Albrecht}
\affiliation{Department of Physics, UC Davis}

\date{\today}

\begin{abstract}

We compare two holographic arguments that impose especially strong
bounds on the amount of inflation. One comes from the de~Sitter Equilibrium
cosmology and the other from the work of Banks and Fischler. We
find that simple versions of these two approaches yield the same bound
on the number of e-foldings.  A careful examination reveals that
while these pictures are similar in spirit, they are not necessarily
identical prescriptions.  We apply the two pictures to specific
cosmologies which expose potentially important differences and which also
demonstrate ways these seemingly simple proposals can be tricky to
implement in practice.

\end{abstract}

\maketitle


\section{Introduction}
There have been a number of attempts to apply notions of holography
towards constraints on inflation. \cite{Albrecht:2011yg,
  ArkaniHamed:2007ky, Banks:2003pt}.  The motivation is simple:
holography implies an encoding of bulk information in correlations on
a boundary, while inflation promotes quantum fluctuations to all
scales and (while it lasts) seems to a provide a fertile source for
new information.  Where these notions come in conflict we can look to
place limits on the amount of inflation allowed; but the strength and
form of the limits will depend strongly on the particular holographic
approach adopted.   

Holographic limits on inflation are of particular interest in the
context of ``eternal
inflation''~\cite{Steinhardt:1982kg,Linde:1982ur,Albrecht:2013lh},
which uses simple extrapolations 
within effective field theory (EFT) to suggest that at some high level inflation
continues forever, giving birth to unbounded numbers of ``pocket
universes''.  Holographic limits may give hints about how a
deeper theory would lead to a breakdown of the EFT, and perhaps
dramatically alter the eternal inflation picture (and perhaps
resolve the notorious measure problems associated with eternal inflation)\cite{Guth:2007ng,Albrecht:2009vr}.

One example of an inflation model limited by holographic arguments is
the de~Sitter Equilibrium (dSE) picture
\cite{Albrecht:2009vr,Albrecht:2011yg, Albrecht:2014eaa}. In that
picture the universe is fundamentally finite, 
with a maximum entropy associated with the asymptotic de~Sitter
horizon at late times.  Because of the finite Hilbert space, the
standard EFT description of inflation will fail if asked to
model a length of inflation producing enough volume to exceed
the universe's maximum information content.  In \cite{Albrecht:2011yg}
this bound manifests as a sharp prediction for
spatial curvature of our universe, as a function of initial bubble
curvature.  In this paper we will show the bounds achieved in dSE are
identical to those found by Banks and Fischler (BF)~\cite{Banks:2003pt}
despite a treatment that incorporates holography differently.
However, we show that specific assumptions chosen by BF
in addition to their maximum-entropy method of deriving a bound on
inflation are needed together to enforce the same geometric principles
used in the 
dSE curvature prediction.  As demonstrated in~\cite{Cai:2003zs}, adjusting
those assumptions to ones more representative of our own universe can
modify the prediction for a maximum number of e-foldings 
of inflation.  Although not immediately apparent, some of those
modifications would result in different physical pictures and
indeed could produce different bounds from the dSE case. 
Each picture has originally been presented in fairly simple
terms, and our work exposes ways the simple definitions appear to be
insufficient to allow for a full implementation in all cases.  This is
how the additional assumptions can become especially important. 
It appears that a geometric interpretation of the holographic principle
along the lines of \cite{Albrecht:2011yg} is useful to clarify 
a number of these issues. 
	
Our paper is organized as follows.  In Sec.~\ref{sec:holo} we
quickly present versions of each picture (BF\footnote{We will repeatedly refer to the result found
in the first half of the paper by Banks and Fischler \cite{Banks:2003pt} as the BF picture.  The project of
our paper is a comparison between the holographic principle underlying
this particular result and the one within the dSE picture, and 
to that purpose we will add specifications, and reflect upon
assumptions and motivations for the ``BF picture''.  These reflections
and modifications are entirely those of this paper's authors and we do
not mean to attribute them to Banks and Fischler.} and dSE) with many 
simplifying cosmological assumptions and approximations. As we show, these
approximations cause the two pictures to converge not only on each
other, but also on cosmologies with past histories similar to ours
(thus suggesting that these ideas are quite relevant to our universe).
In Secs.~\ref{sec:bounds} and \ref{sec:same} we study this apparent correspondence.  In
Sec.~\ref{sec:assume} we begin to tease apart the two pictures by examining
the assumptions required for each.  In the first step we simply ask
what additional clarifications are needed in each picture for
applications to more
realistic and specific cosmologies.  It is generally possible to
ensure that the BF and dSE pictures return the same results, but  
the process of picking the ``correct" choice of assumptions to match
the two begins to look ad hoc.
Finally we examine the two
principles with cosmologies that are  parametrically connected to ours
but substantially altered.  In this manner we can better expose conceptual differences between the two approaches  
when pushed away from their convergence near our own relatively simple
cosmological history.

\section{Holography in dSE and BF \label{sec:holo}}
\subsection{General discussion}
In both the dSE and BF pictures, the future of our universe is
asymptotically de~Sitter, with a fundamental cosmological constant
$\Lambda$.  In both of these pictures the entropy $3\pi/\Lambda$  
associated with the de~Sitter horizon represents the finite amount of
information associated with the entire universe.  In such a picture it
is expected that physics can describe semiclassical spacetimes with a
maximum of one horizon volume (although different observers can
observe different realizations of such a volume by swapping out information encoded
non-locally at the de~Sitter horizon with the interior). 

Though the horizon entropy is important in the setup of dSE cosmology,
Banks and Fischler do not use it in their calculation of the bound on
the number of e-foldings. Instead, the universe is
modeled as a fluid-filled cavity the size of the apparent horizon, and
the entropy within that cavity is used to obtain the bound.  

It is important to distinguish between these holographic principles and
other variants that physically differ and result in different (or no)
bounds on the amount of inflation.  For example, the covariant Bousso
entropy bound~\cite{Bousso:1999xy} is formulated on the past light
cone of an observer.  Placing such bounds on the past light cone does
not restrict the number of e-foldings of inflation~\cite{Lowe:2004zs}. In a similar manner Kaloper {\it et al.}~\cite{Kaloper:2004gp} interpret the
BF bound as only placing limitations on the number of e-foldings of inflation
that will ever be observable. In contrast, in the dSE picture the entire
universe is eventually observable so there is no distinction to be
made between observable and total e-foldings.


Numerous authors have proposed bounds on inflation under a variety of
assumptions. For example, Arkani-Hamed {\it et al.} found a much less stringent bound of
$N_{total} < S$ (where $S$ is the entropy of the final asymptotic de
Sitter space) by demanding non-eternal inflation~\cite{ArkaniHamed:2007ky} . Albrecht {\it et al.} also put
forward another holographic inflation bound by using the slowly changing apparent horizon to estimate the 
entropy of metric fluctuations expelled during the slow roll period of inflation~\cite{Albrecht:2002xs}. Bousso's D-bound originates from
the requirement that entropy not decrease during the transition to empty de
Sitter space and works by positing that the entropy gained from the
increased horizon area must exceed that of the matter entropy that
was lost~\cite{Bousso:2000nf}.
These bounds each emerge from fundamentally
different holographic principles and we feel each is 
interesting in its own right.  Here we restrict our attention only
to the bounds from the dSE and BF pictures because they seem to admit
direct comparison. 


Finally, for a given holographic principle it is important to
distinguish between an absolute bound on the length of inflation for
any universe (e.g. allowing for the most extreme variations in
reheating, matter fraction, etc. consistent with some set of
cosmological assumptions), and a bound on inflation for a universe
consistent with the one we observe.  The first bound is of more
interest for exploring a multiverse of cosmologies, either to
understand the allowable regions within a theory or for making
predictions within a multiverse.  Constructing the second type of
bound is more relevant for direct connections to observations.
For now we will restrict ourselves to simplifying
assumptions relevant for our universe, but later (in
Sec.~\ref{sec:assume}) we will expand our focus to a broader range of
cosmologies. 

\subsection{dSE bound basics}
Under the de~Sitter Equilibrium picture's assumption of a finite universe whose size 
is set by a fundamental cosmological constant~\cite{Albrecht:2009vr}, an observer near the universe's
final approach towards de~Sitter space should be able to see
essentially all that there is in the universe.  Requiring the past
horizon of such a ``maximal observer" to contain all scales produced
by inflation puts a bound on the maximal length of inflation, which
could otherwise generate structure that grew to physical scales
arbitrarily larger than the universe's size by the time of the maximal
observer.

\subsection{Banks Fischler bound basics}
Banks and Fischler \cite{Banks:2003pt} follow the entropy to examine how
an ultimate size limit for the universe restricts the length of
inflation.  They note a number of results restricting the maximum
entropy within a sphere for a non-collapsing fluid with a given
equation of state, and then demand that the ultimate entropy of the
universe be no larger than that limit calculated for a sphere of the
universe's ultimate size, for the appropriate fluid.  As an initial
patch inflates its volume increases, so the
limit on the length of inflation arises by requiring that upon 
reheating the total entropy of the entire inflated region does not
exceed this calculated fundamental limit.  

\subsection{The connection between the two pictures}
The BF and dSE pictures use holography-inspired principles that are very similar
and amount to restricting the ultimate radius of the universe to
approximately the de Sitter radius.  In both pictures, by the onset of
de~Sitter domination we expect an observer to be able in principle to
observe everything produced during inflation, rather than allowing some matter
to remain forever out of reach.  From these two facts it
appears as though the two approaches are guaranteed to deliver
essentially the same bounds on inflation, and to a certain degree this
is true.  However, as we will see in Sec.~\ref{sec:assume}, the pictures are not necessarily physically
identical, and can only be made so with the buttressing of enough
simplifying assumptions to force the two pictures to describe
identical scenarios.  The logical differences between the two remain
of interest and investigating what is required to bring the two into
alignment helps to clarify both pictures.

\subsection{Counting e-foldings in dSE: A geometrical picture}
\begin{figure}[h]
\begin{center}
\includegraphics[scale = 0.78]{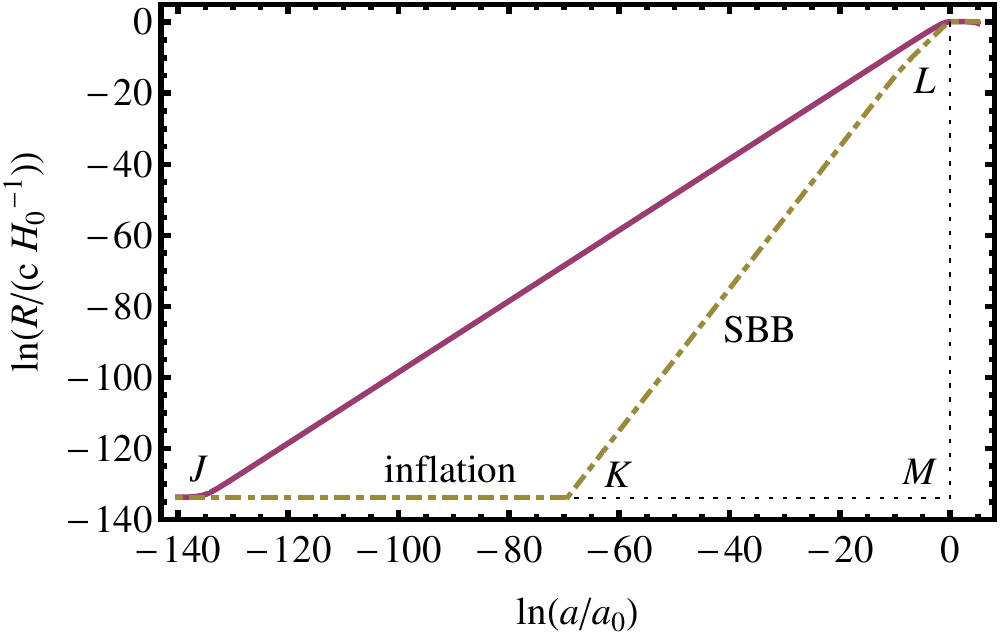}
\caption{The evolution of various length scales with scale factor $a$
  plotted during inflation and the subsequent standard big bang
  expansion (SBB) up   to the dS scale for a simple model of our
  universe. Plotted are the Hubble length
(dot-dashed) and the past horizon $h_P$ of an event late in the de
  Sitter era (solid). The radiation-matter transition occurs at
  $ln(a/a_0) = -8.1$ 
  and shows up as a very slight kink in the dot-dashed curve. Markers
  $J$,$K$,$L$ and $M$ aid the discussion in the text.\label{fig:RH_vs_a}} 
\end{center}
\end{figure}

%
Figure \ref{fig:RH_vs_a} shows the evolution of the Hubble radius
$R_H \equiv c H^{-1}$. Also shown is $h_P$, the past horizon of an
event late in the cosmological constant dominated
regime. Specifically, if the event is the observation of a photon which
travels freely before the observation, $h_P$ is the 
distance between the photon and the observer prior to observation. 
Owing to the formation of an event horizon as we
approach a de~Sitter background, the past horizons of events at any
time in the de~Sitter regime are much the same. Any of these
observers can see essentially as much volume as any observer ever will.  Even today
($a=a_0$) we are not too far off from being such ``maximal observers'', since
$\Lambda$ is already quite dominant. The
cosmological parameters for the curves shown match our universe with
reasonably fast reheating assumed.  

The formulation of dSE cosmology~\cite{Albrecht:2011yg} requires the
entire universe to be contained within the past horizon of such a
maximal observer.  If our classical description of inflation begins
earlier than shown
in Fig.~\ref{fig:RH_vs_a}, there would be scales
which never re-entered the horizon of this maximal observer, and which
would represent physical scales larger than the finite
size ($\sim \Lambda^{-1/2}$) assumed for the universe.  This requirement places
a limit on the amount of inflation allowed in dSE.   

The form of the dSE constraint leads to a particularly simple
connection with the cosmic curvature. Through
most of the universe's history, the past horizon scale evolves 
$\sim a$. Because the spatial curvature radius also evolves $\sim a$,
one can make a sharp prediction for today's measured spatial
curvature density $\Omega_k$ in terms of the spatial curvature of the
bubble that began inflation~\cite{Albrecht:2011yg}.  
Due to the geometric nature of this picture the details of
reheating and the subsequent evolution of the universe can modify the
length of inflation, but they do not
change the prediction for the curvature. Thus curvature is a more
robust reflection of the dSE bound than number of e-foldings.
Nonetheless, one can apply this geometric framework to derive bounds
on the length of inflation as well. 


The geometric notions in dSE allow us to calculate calculate the bound
on the length of inflation in the dSE picture because the curvature
prediction amounts to considering the geometry of a plot such as
Fig.~\ref{fig:RH_vs_a}. 
For simplicity we'll approximate the evolution of the physical photon
distance by two line segments (replacing the smooth transition region
near $a=a_0$ with a sharp corner).  For the segment representing past evolution, the distance
evolution is proportional to $a$, so its slope is unity in
Fig.~\ref{fig:RH_vs_a}.  We will also assume that during inflation the
Hubble parameter $H_I$ is approximately constant and the subsequent
reheating is rapid.  After reheating we treat the universe as a perfect
fluid with a single equation of state $p = w\rho$ right up to the
beginning of the dS stage (which sets in with $H = H_{\Lambda}$ at scale factor 
$a_{\Lambda}$).  Thus we are also representing the $R_H$ curve in
Fig.~\ref{fig:RH_vs_a} with straight line segments meeting at sharp
corners.  Since most of the logarithmic range of $a$ is in the radiation
era (vs matter era), and the slope of the $R_H$ curve during the
matter era is not that different from the radiation case, one can
achieve a good approximation to Fig.~\ref{fig:RH_vs_a} by taking $w=1/3$
in our linear approximation. 

With these simplifications and the dSE assumption fixing the beginning
of inflation on the past horizon line, the increase in the Hubble
radius $H^{-1}$ from inflation to dS must equal the increase in the
scale factor over the same time:
$\ln\frac{H_{\Lambda}^{-1}}{H_I^{-1}} = \ln \frac{a_{\Lambda}}{a_I}$. On Fig.~\ref{fig:RH_vs_a} this corresponds to setting the lengths ${\rm \it JM = LM}$.

To find the number of e-foldings for inflation ($\rm \it JK$), we simply
start with the e-fold increase in the
Hubble length ($\rm \it JM$) and subtract off the e-folds of $a$ ``eaten up" during
the standard big bang expansion (SBB) of the fluid to the dS scale ($\rm \it KM$).  Since $H^{-1}
\sim \rho^{-1/2}$ and $\rho \sim a^{-3(1+w)}$, $H^{-1} \sim
a^{3(1+w)/2}$, which gives us the slope during SBB.  Thus the
increase of $a$ during SBB is given by $\frac{2}{3(1+w)}
\ln\frac{H_I}{H_{\Lambda}}$, and the number of e-foldings $N_e$ of
inflation is  
\begin{align} \label{dSE_bound}
	N_e &= {\rm \it JM - KM} \\
	N_e &= \ln\frac{H_I}{H_{\Lambda}} (1 - \frac{2}{3(1+w)}) \\
		&= f\ln\frac{H_I}{H_{\Lambda}}, \quad \quad f \equiv \frac{1}{3}(3 - \frac{2}{(1+w)})
\end{align}
%
%
%

In the case of radiation ($w = 1/3$), $f = 1/2$ and the universe's
history is evenly split: the magnitude of the universe's expansion
during inflation equals that from reheating through the start of
$\Lambda$ domination.  For GUT-scale $H_I \approx 10^{40}$ s$^{-1}$ and the
observed $H_{\Lambda} \approx 10^{-18}$ s$^{-1}$, that gives $N_e =
\frac{1}{2} \ln\frac{H_I}{H_{\Lambda}} \approx 67$
e-foldings. Decreasing $H_I$ will increase $R_H$ during inflation,
raising the horizontal line segment in Fig.~\ref{fig:RH_vs_a}, and thus
decreasing the bound on $N_e$, assuming the other elements of the
calculation are held fixed. 
%


\section{Entropy bounds for fluids systems \label{sec:bounds}}
\subsection{Overview}
Banks and Fischler's holographic approach involves counting the
entropy in a cavity the size of the apparent horizon. With this method
of entropy counting, putting a bound on inflation involves putting a
bound on the entropy within this cavity. Thus we will find it useful
to discuss bounds on the amount of entropy that can be contained
within a cavity of size $R$ without it collapsing due to
gravity. Banks and Fischler examine a number of approaches to bound
the entropy of a fluid system. In each case the scaling obtained for
fluids with equation of state $p = w\rho$ is of the form 
\begin{align}
\label{entropy_relation}
S \le \beta \left(\frac{R}{l_p}\right)^{3-\frac{2}{1+w}},
\end{align}
where $l_p$ is the Planck length and $\beta$ is a constant
determined by the thermodynamic relation for the entropy density of
the fluid, discussed in more detail in Sec.~\ref{sec:thermo}. For
radiation, $\beta=O(1)$ in these units  
one Planck volume should contain at most roughly one unit of
entropy. The common form (Eqn.~\ref{entropy_relation}) for these
results formed the basis of the derivation for the BF inflation
bound. Banks and Fischler derived their bound in a simple
cosmology and we next reproduce their argument in the remainder of
this section.

\subsection{Critical (flat) FRW universes without dark energy \label{sec:flatfrw}}
We write the Friedmann equation as 
\begin{align}
H^2 = \frac{8\pi G}{3} \rho_{tot} - \frac{k}{a^2}.
\end{align}
Using the thermodynamic relation for the entropy
density 
\begin{align}
 \sigma = \beta \rho^{\frac{1}{1+w}} \label{eq:thermo}
 \end{align}
of a fluid with equation of state $p = w \rho$, we can calculate the
total entropy contained within a Hubble volume $H^{-3}$ for a flat ($k
= 0$) universe: 
\begin{align}
S = H^{-3} \sigma = \beta H^{-3}\rho^{\frac{1}{1+w}} = \beta H^{-3 + \frac{2}{1+w}}
\end{align}
The factor $\beta$ is at most $O(1)$ for a single species (for $\sigma$ and $\rho$ expressed in Planck units) but can be
significantly smaller, as discussed in Sec.~\ref{sec:thermo}.

\subsection{Universes with $\Lambda > 0$}
Similar results exist for certain cases in universes with a
cosmological constant~\cite{Banks:2003pt}.  
Fischler et al.~\cite{Fischler:2003xq} consider how much entropy can
be stuffed into a region without it collapsing by relating the
energy density to the entropy density using the thermodynamic
relation~\eqref{eq:thermo}, and then solving the Friedmann equations. 
They point out that for static solutions in matter-dominated ($w=0$)
universes, there is a 
metric sign change at precisely the same 
entropy bound given by Eqn.~\eqref{entropy_relation}.  
Fischler et al. show that violating this bound in the collapsing phase
of a universe with positive $\Lambda$ causes a big
crunch~\cite{Fischler:2003xq}.    

In both dSE and the BF approach the finite size of the universe places
an upper limit on the radius of a fluid-filled sphere.  This maximum
radius would then imply a maximum theoretical entropy for such a
fluid-filled universe.  The repeated appearance of the
relation~\eqref{entropy_relation} encouraged BF~\cite{Banks:2003pt} to
ask what limitations on inflation could result if one demands the
global entropy produced during inflation to remain less than this
entropy bound evaluated at the maximal radius $\sqrt{3/\Lambda}$. Our
paper seeks to compare the resulting bound on inflation with related
results within dSE. 

\subsection{BF e-fold counting}
Banks and Fischler \cite{Banks:2003pt} arrived at a formula for
e-foldings identical to Eqn.~\ref{dSE_bound} in the first half of their paper.\footnote{With the
  addition of a ``holographic gauge" condition, they later produce a
  different estimate.   As dSE cosmology does not incorporate an
  analogous condition, we will restrict our comparison to the first
  estimate.} 
BF adopt a holographically inspired view that functionally bounds
space within a cavity of radius $\Lambda^{-\frac{1}{2}}$. As discussed
in Sec.~\ref{sec:flatfrw}, filling a cavity of this size with a fluid
of equation of state $p=\rho w$ results in a maximum entropy $S \lesssim
R^{3 - \frac{2}{1+w}}$.  Their approach is to assume that the inflaton
reheats into a fluid with state parameter $w$ at a density $\rho_i
\sim H_I^2$, with an entropy density $\sigma_i = \beta
\rho_i^{\frac{1}{1+w}}$.  At the reheating time there are $e^{3N_e}$ Hubble patches.  The goal is to ensure that if
we sum all the entropy in these patches, we do not exceed the limit  
\begin{align}
S_{max} \sim H_{\Lambda}^{-(3 - \frac{2}{1+w})} \label{dSE_result}
\end{align}
 for the asymptotic apparent horizon  $\sim H_{\Lambda}^{-1}$.
	
	The entropy $S_i$ in a single Hubble volume after reheating is 
\begin{align}
	S_i &= H_I^{-3} \sigma_i \\
		&= H_I^{-3} \beta \rho_i^{\frac{1}{1+w}} \\
		&\sim H_I^{-(3- \frac{2}{1+w})}
\end{align}
where $H_I$ is the value of the Hubble constant at the end of
inflation and (as in~\cite{Banks:2003pt}) we have suppressed
pre-factors such as $\beta$.  Thus the
entire volume of $e^{3N_e}$ Hubble patches must obey 
\begin{align}
	e^{3N_e} S_i &\le S_{max} \\
	e^{3N_e} H_I^{-(3- \frac{2}{1+w})} &\le  H_{\Lambda}^{-(3 -
          \frac{2}{1+w})} \\
\end{align}
giving
\begin{align}
	N_e &\le \frac{1}{3}(3-\frac{2}{1+w})\ln \frac{H_I}{H_{\Lambda}}
\end{align}
which is identical to the dSE case (Eqn. \ref{dSE_bound}).

\section{How the BF and dSE bounds wind up the same \label{sec:same}}
We would like to relate the BF result as closely as possible to the
dSE result.  We can phrase the dSE bound most simply as the
requirement that the increase in physical volume $a^3$ from the beginning of inflation
to the beginning of the dS era equals the increase in Hubble volume
over the same period. At first glance the BF bound does not depend on
such geometric notions; it merely demands the global entropy produced
at reheating be no more than the maximum allowed for a fluid that can
fill a cavity the size of the de~Sitter horizon without collapse. 

In order to connect the BF picture and the
geometric ideas from dSE one can follow a comoving 
region of space through the evolution of the universe. We focus on the
region C bounded by the apparent horizon when inflation
starts. (The dashed line in Fig.~\ref{fig:RH_vs_a_zoom_box} shows the size of region C.) 
During inflation, C expands exponentially, but the size of the apparent horizon A
stays the same. After reheating the resulting
radiation dominated universe C expands too, but the region that is
contained within the apparent horizon A expands faster. This is also
the case during matter 
domination. When we reach cosmological constant domination, both A and
C have become very large and we put a bound on inflation by requiring
A to be contained within C at all times.\footnote{Imposing this
  requirement at all times is implicit in the BF analysis for simple
  cosmologies but in general represents an additional assumption.} Since C describes a comoving
region the entropy in this region  is
conserved assuming adiabatic evolution. 

Because the underlying restriction on the size of the universe is the
same in both approaches, it may seem that the bounds on inflation are
destined to be identical.  But while this restriction is explicit in the
derivation for the dSE case, it does not appear directly in the BF
derivation. The assumed adiabatic expansion of the universe is naturally tracked by comoving volumes of constant
entropy, and it is this translation to the language of comoving volumes that allows contact with the
geometric dSE picture. 
As we will discuss in Sec.~\ref{sec:reexamine}, adiabaticity is only one of several assumptions required
for the simplest version of BF to agree with dSE.

We will show that relaxing the simplifying
assumptions behind the BF bound can lead to a physically different
scenario with different limits on the length of inflation or
predictions for curvature.  Only by explicitly requiring a geometric
statement of the holographic principle as part of the BF analysis do
we ensure that it is actually imposing the same constraint as in the
dSE analysis. 

In the next section, we will expand the investigation into the
assumptions required in order to match the BF and dSE pictures. We
will look at more complicated examples consistent with our universe
and also more general cases.  Considering these cases will reveal  some
issues that arise when pursuing a rigorous holographic bound valid for
all cosmologies.  

\section{Re-examining assumptions for the BF and dSE bounds \label{sec:assume} \label{sec:reexamine}}
\subsection{Overview}

Here we give a more detailed account of key assumptions and
simplifications that go into the BF and dSE 
bounds.  This will help us examine how these assumptions relate to
the equivalence (or not) of the BF and dSE bounds. First we give a
descriptive list of these assumptions and then elaborate on each one
in separate subsections. 

\textbf{Horizons:} Both BF and dSE involve identifying horizons,
  but there are subtleties in using these horizons that need to be
  understood in order to make a sharp comparison. 

{\textbf{Prefactor:} The simple thermodynamic entropy scaling relation
  of Eqn.~\eqref{eq:thermo} has a pre-factor that requires scrutiny.} 

{\textbf{Net equation of state:} The BF picture relies on a fluid with a 
	single equation of state throughout the cosmic evolution.
        There are choices involved in defining a  
	single effective equation of state for a realistic universe
        composed of multiple fluids with different equations of state.} 

{\textbf{Adiabaticity of the fluid:} We also need to account for possibly substantial ordinary entropy production
  that does not necessarily change the cosmological equation of state, such as particle decays or stellar processes.}

{\textbf{Black holes:} Universes do form black holes, and we need to examine the BF approach of
	excluding black hole entropy.}

{\textbf{TOV equation:} We will examine solutions of
  maximum entropy permissible within a non-collapsing universe
  with cosmological constant, represented by the
  Tolman-Oppenheimer-Volkoff equation. This will allow us to
  generalize beyond the homogeneous.} 

{\textbf{Alternate cosmologies:} We will explore how well the two pictures can describe non-collapsing 
  cosmologies that exhibit a loitering period of slow
  expansion, allowing observation of an arbitrarily large volume of
  the universe.}

A careful look at each of these issues will help us get
a better understanding of the challenges involved in formulating
such types of holographic bounds, both for our universe, and in
general. 

\subsection{Choice of horizons}
\begin{figure}[h]
\begin{center}
\includegraphics[scale = 0.78]{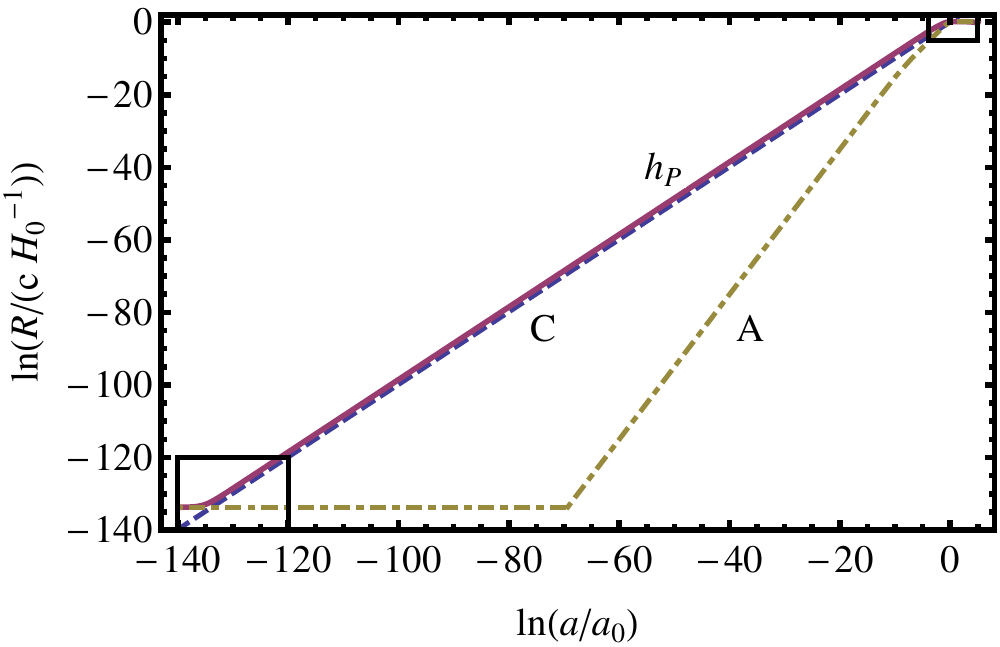}
\caption{Evolution of length scales $R$ with scale factor $a$. Plotted are the
  size of the apparent horizon A (dot-dashed) and the past horizon $h_P$ of an event late in the de~Sitter era (solid). The dashed line C tracks the
  comoving volume of space with the maximal entropy that can be
  contained in the fluid cavity of BF assuming adiabaticity. Zoomed
  versions of the boxed
  regions are shown in Figs.~\ref{fig:RH_vs_a_late} and 
\ref{fig:RH_vs_a_early}. (In these zoomed figures the differences
between $C$ and
$h_p$ appear clearly). \label{fig:RH_vs_a_zoom_box}}
\end{center}
\end{figure}

Banks and Fischler count the entropy at the exit of inflation by modeling the universe as a fluid-filled sphere with a size equivalent to the apparent horizon~\cite{Banks:2003pt}.
In dSE, the past horizon is used, specifically
\begin{align}
h_P(a_1) \equiv a_1 \int_{a_1}^{a_\Lambda} \frac{d a}{a^2 H}.
\end{align}
Figure~\ref{fig:RH_vs_a_zoom_box} shows the evolution of both these horizons in a way which allows
us to visually express the bounds on the number of
e-foldings of inflation in both the dSE and BF pictures. 

\begin{figure}
\begin{center}
\includegraphics[scale = 0.78]{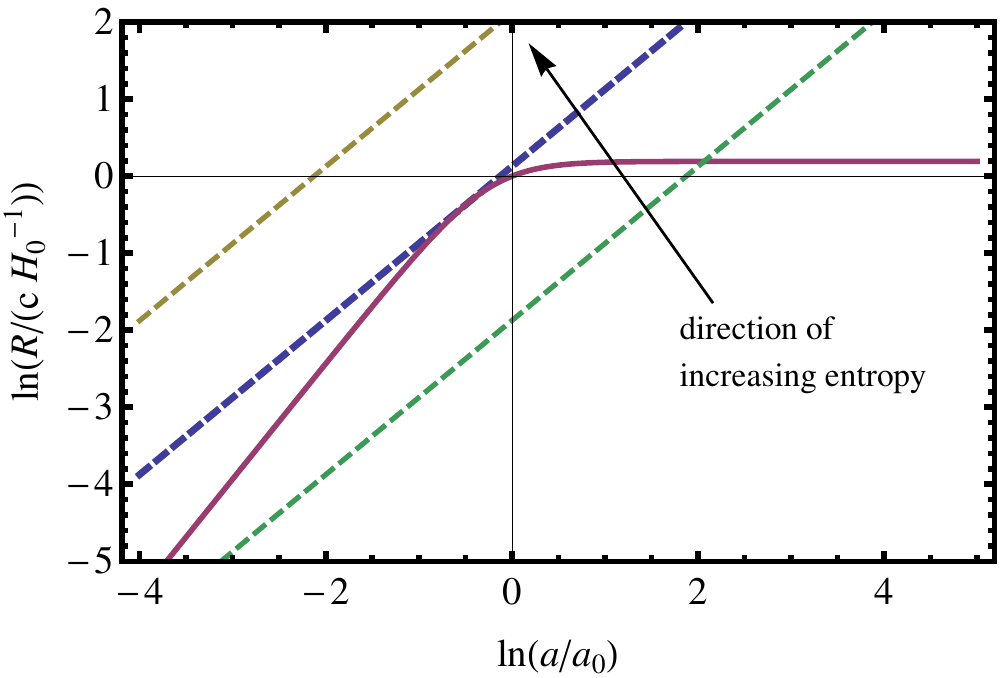}
\caption{Evolution of various length scales vs. scale factor
  $a$. Plotted are the   size of the apparent horizon (solid) and
  a few lines of constant entropy (dashed). The comoving volume of
  space with the maximal entropy that can be 
  contained in the fluid cavity of BF assuming adiabaticity is the
  thick central dashed line tangent to the solid
  curve. \label{fig:entropycontours}} 
\end{center}
\end{figure}

We evaluate the BF bound by requiring that the entropy
always be less than the maximal entropy contained within the
fluid-filled cavity the size of the apparent horizon.  Assuming
adiabaticity 
comoving regions contain constant entropy and can be
represented by $R \propto a$ lines in
Fig.~\ref{fig:entropycontours}. The heavy dashed line in
Fig.~\ref{fig:entropycontours} (also shown as a dashed line in
Fig.~\ref{fig:RH_vs_a_zoom_box}) indicates the largest comoving region 
ever contained within the apparent horizon (solid curve), and is thus
the natural focus of the BF analysis.  
To calculate the BF bound on the number of
e-foldings geometrically we follow the dashed line in
Fig.~\ref{fig:RH_vs_a_zoom_box} back to where it intersects the apparent
horizon in the inflationary epoch, which marks the earliest allowed
start to inflation in the BF picture. We read off the bound on the
number of e-foldings as the horizontal distance between this
intersection and the end of inflation.  

\begin{figure}
\begin{center}
\includegraphics[scale = 0.78]{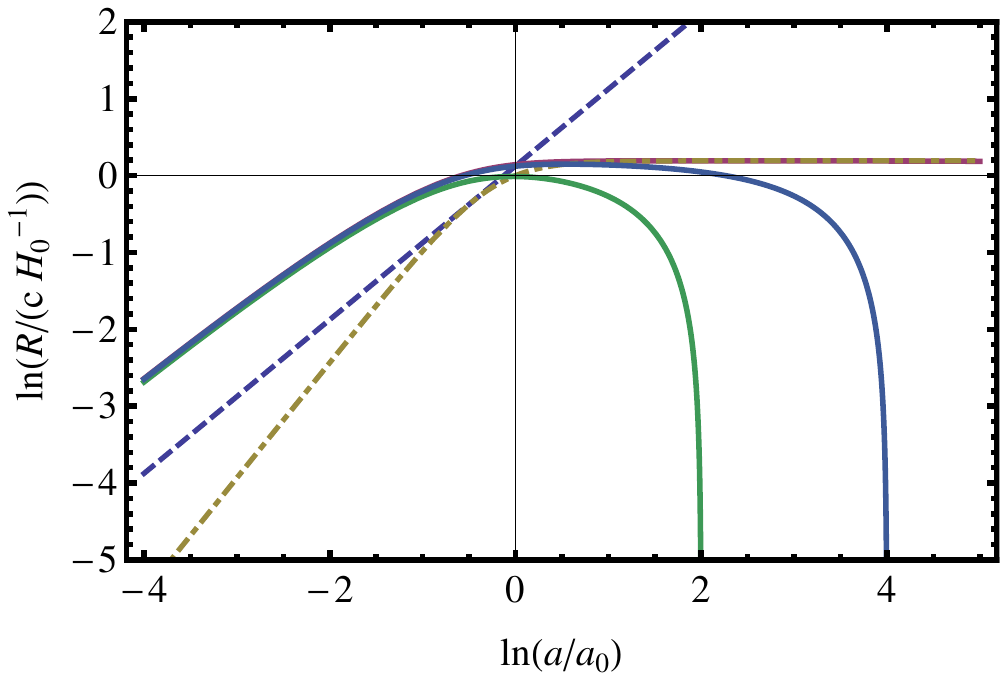}
\caption{Close-up of Fig.~\ref{fig:RH_vs_a_zoom_box} near the
  era of cosmological constant domination. The dot-dashed curve is the 
  apparent horizon and the solid curves show past horizons for events
  at a few different times in the de~Sitter era.  The feature
  that these past horizons all approach each other at early times is due
  their being defined relative to events in the de~Sitter era.
 The dashed line tracks the comoving volume of space
  with the maximal entropy that can be contained in the fluid cavity
  of BF assuming adiabaticity. \label{fig:RH_vs_a_late}} 
\end{center}
\end{figure}
\begin{figure}
\begin{center}
\includegraphics[scale = 0.78]{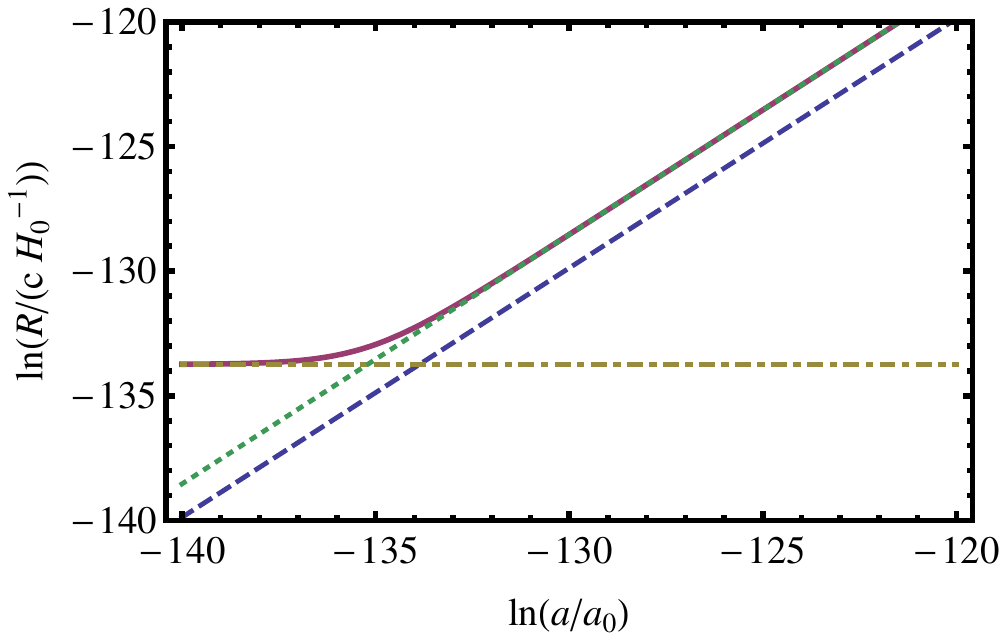}
\caption{Close-up of Fig.~\ref{fig:RH_vs_a_zoom_box} plotted near the
  beginning of inflation. Plotted are the size of the apparent horizon
  (dot-dashed) and the same $h_P$ curves shown in
  Figs.~\ref{fig:RH_vs_a_zoom_box} and~\ref{fig:RH_vs_a_late}
  (solid). The dashed line tracks the 
  comoving volume of space with the maximal entropy that can be
  contained in the fluid cavity of BF assuming adiabaticity. The
  dotted line shows the approximation $h_P \propto a$. The
  difference between the dotted and dashed lines corresponds to 
  an extra 1.3 e-foldings of inflation more in the dSE picture than in
  the BF picture. \label{fig:RH_vs_a_early}} 
\end{center}
\end{figure}

In the dSE picture the universe is bounded by the past horizon which
can be well approximated by $h_P 
\propto a$ (the dotted line in Fig.~\ref{fig:RH_vs_a_early}) for most
of the evolution of the universe. We can see how 
this approximation can be used to picture the dSE bound in a similar
way to how we just described the BF bound: The earliest allowed start to inflation in the dSE picture is given by where the dotted line in Fig.~\ref{fig:RH_vs_a_early} intersects the apparent horizon in the inflationary epoch.


The $h_P \propto a$ approximation breaks down at both early and late
times. In the late time era of cosmological constant domination
(shown in detail in Fig.~\ref{fig:RH_vs_a_late}), looking back
from a later time does not noticeably change the past horizon. This is
a nice feature since it means that it does not matter when we choose
to observe as long as it is during cosmological constant
domination. However the breakdown of the approximation at early times
(detailed in Fig.~\ref{fig:RH_vs_a_early}) presents some problems for
constructing an accurate prescription for a 
bound on inflation. The past horizon approaches the apparent horizon
at early times but never crosses it, seemingly implying that there is
no bound. However, in the dSE picture we are inclined to say that the
deviation from the approximation at early times occurs at a time where
we expect new physics and the breakdown of the effective field
theory (EFT). Without a clear picture of what lies beyond the EFT we
simply use the intersection of dotted line (the $R \propto a$
extrapolation of the past horizon) with the dot-dashed curve (the
apparent horizon) to indicate the effective start to inflation.


The distinction between the two horizons seems like a small technical
difference, but as we will see in Sec.~\ref{sec:loiter}, the
distinction can become problematic for universes with large curvature
or periods of slow expansion. 

%
%





\subsection{Thermodynamic relation for entropy density \label{sec:thermo}}
The expression for the entropy density of an adiabatically expanding fluid, $\sigma = \beta \rho^{\frac{1}{1+w}}$, comes from statistical mechanics. We would like to evaluate the proportionality factor to make this relation more precise. For a particle species in thermodynamic equilibrium,
\begin{align}
\rho &= \frac{g}{(2 \pi)^3} \int E \frac{d^3 p}{e^{(E-\mu)/T} \pm 1} \\
P &= \frac{g}{(2 \pi)^3} \int \frac{p^2}{3 E} \frac{d^3 p}{e^{(E-\mu)/T} \pm 1} \\
\sigma  &=  \frac{\rho+ P}{T},
\end{align}
where $\rho$ is the energy density, $P$ is the pressure, $\sigma$ is the entropy density, $g = g_B + \frac{7}{8} g_F$ is the total effective number of
internal degrees of freedom, and $g_B$ and $g_F$ are the number of
spin states for bosons and fermions respectively. 
For example, in the case of relativistic particles,
\begin{align}
\rho &= g \frac{\pi^2}{30}\frac{(k T)^4}{(\hbar c)^3} \\
\sigma  &=  g k \frac{2 \pi^2}{45}\left(\frac{k T}{\hbar c}\right)^3 \\
\sigma &\approx 1.0098 g^{\frac{1}{4}} k (\hbar c)^{-\frac{3}{4}} \rho^{\frac{3}{4}} \label{eq:entropyrad},
\end{align}
and we find that the pre-factor in this case is indeed of order one (as long as the number of internal degrees of freedom is not extraordinarily large). 
If we have a fluid of relativistic particles at the Planck density in
thermal equilibrium, it will be at roughly Planck temperature and if
it is contained in a Planck volume, we see from the expression above
that it will have of order one unit of entropy as expected. 

We will also look at what happens during the early universe in a
typical model where a thermal relic decouples non-relativistically. A
typical scenario is that the inflaton reheats into relativistic
particles, some of which are massive and will eventually cool and
become non-relativistic. The non-relativistic matter will then typically
freeze out and as it does so, asymptote to a constant comoving number
density. 

\begin{figure}
\begin{center}
\includegraphics[scale = 0.78]{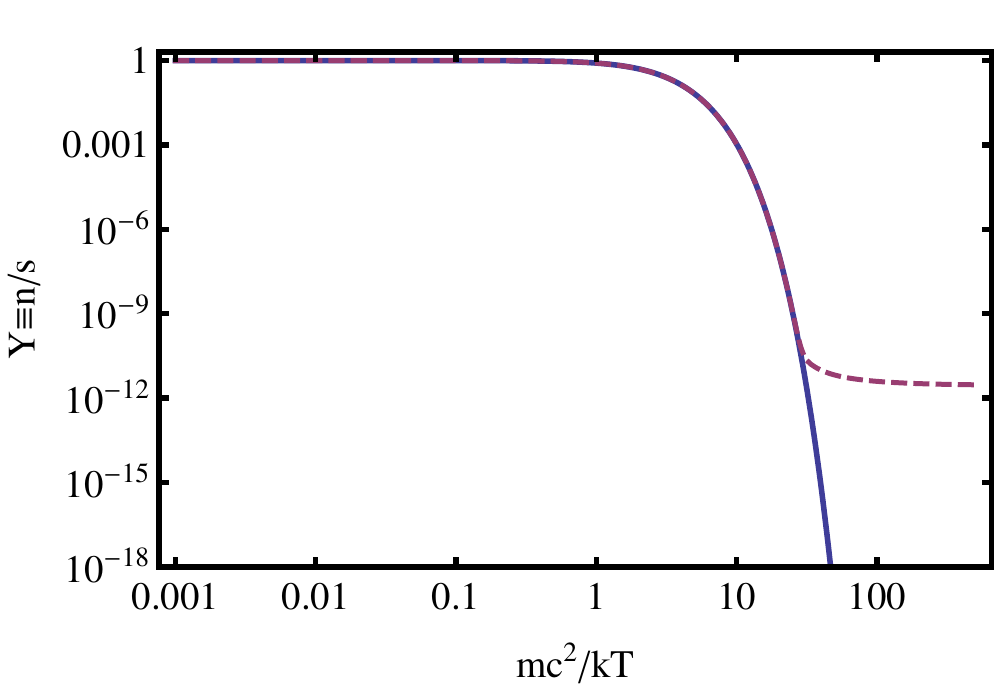}
\caption{The generic behavior of entropy density as a function of time
  during freeze-out. As the universe cools, the number of particles
  drops until the particles cannot find each other to annihilate and
  freeze out. The solid line indicates the equilibrium abundance,
  while the dotted line is the actual particle
  abundance. \label{fig:relic}} 
\end{center}
\label{relic}
\end{figure}

In general, non-relativistic matter will possess less entropy than
radiation.  If we trace the matter's history back to a time when its
temperature was higher than its mass, it had the same entropy per
degree of freedom as the relativistic case.  However during its
transition to a non-relativistic fluid the matter's comoving number
density (and entropy density) exhibits a characteristic drop as it
cools before asymptoting to a constant value again during freeze-out
(Fig.~\ref{fig:relic}). The entropy from the annihilation of particles
is deposited in the radiation. The resulting pre-factor $\beta$ can be
many orders of magnitude smaller than for relativistic matter.

The concept of thermal wavelength is a good way to see the proportionality factors. A fluid in a box the size of the thermal wavelength (not a box of Planck volume) should have entropy of order one. For the same energy density, matter will always have a larger thermal wavelength than radiation.
%


\subsection{Equation of state}
A cosmology like our own universe is not dominated by a single fluid equation of state from reheating
until cosmological constant domination.  Even in the simplest one-component case, the transition to an asymptotic
de~Sitter state modifies the effective equation of state, rounding off the sharp corners in pictures like Fig.~\ref{fig:RH_vs_a}
and adjusting the crudest estimates for inflation bounds. For example, the simple entropy scaling of Eqn.~\eqref{eq:thermo} will begin to 
fail around the time the dominant fluid energy density drops to near $\rho_\Lambda$ (see Fig.~\ref{fig:RH_vs_a_late}).
 In addition, the important transition from radiation
to matter domination in our universe's history sits somewhat uncomfortably with the one fluid model of BF.  

Perhaps the simplest resolution is 
to define an effective equation of state for the entire universe's history 
(in effect drawing the straight line ${\rm \it JL}$ on Fig.~\ref{fig:RH_vs_a}) 
and then proceed with the BF formulation.  But there is something contrived about this approach.  It requires supplying the
entire subsequent evolution of the universe as an input to a calculation formally made at the end of reheating.  
In fact the motivation for this choice is an attempt to keep the BF
picture in compliance with a more geometric approach to the
holographic principle, along the lines of the dSE picture.  If we
insisted on calculating the BF bound in our universe using the
radiation fluid that dominated after reheating and through most of the
expansion history of the universe, we would calculate a larger maximum
number of e-foldings of inflation than if we input the actual
effective `mixed' equation of
state.\footnote{Following~\cite{Fischler:2003xq,Piao:2006ye} we note
  that the change from radiation to matter domination is the reason
  the CMB entropy is less than the maximal BF bound for radiation. We
  can estimate the entropy density of the CMB radiation by using the
  value of $\rho_r$ in Eqn.~\eqref{eq:entropyrad}. Since $\rho \propto
  H_\Lambda^2 = H_0^2 \Omega_\Lambda $ in 
the BF bound while $\rho_r \propto H_0^2 \Omega_r $, the actual
CMB has less entropy than the BF bound by a factor of
$\left(\frac{\Omega_\Lambda}{\Omega_r}\right)^{3/4}$.}
In Sec.~\ref{sec:adiabaticity}
we will see other problematic examples involving changing fluid
constituents.  It should be pointed out that there is not an immediate
``correct'' choice  
for equation of state; no matter how we choose we are forced to decide which clarifications of BF seem most reasonable (or least unappealing).  This is a pattern we will see again.

\subsection{Adiabaticity assumptions} \label{sec:adiabaticity}
Our own universe's early evolution was well approximated as adiabatic, 
but even with the exclusion of purely gravitational entropy its subsequent history
included substantial non-adiabaticity.  And there is no requirement that similar cosmologies be even as approximately entropy-conserving as ours, as it is no great theoretical challenge to come up with mechanisms to increase entropy. 
Moreover, the BF bound was derived under the principle of maximizing entropy that
could be packed into a sphere without collapsing.  One might even wonder
whether the maximal entropy non-collapsing solutions indeed 
are the uniform density solutions assumed so far.  (We 
will further discuss this in Sec.~\ref{sec:tov} and conclude that the
homogeneity assumption is OK.)  If
the universe reheats into a state which does not have maximal entropy
(a realistic case), we are forced to decide among multiple interpretations of
the BF bound.  A treatment of increasing entropy in the
Banks and Fischler picture could alter the inflation bound in either
direction, depending on how one modifies the BF procedure.   

One way to characterize adiabaticity (or its lack) during a cosmology dominated
by a single fluid is simply by using the pre-factor $\beta$ in the
expression $\sigma = \beta \rho^{\frac{1}{1+w}}$.  In the BF picture
for a universe with a fixed effective 
equation of state but substantial non-adiabaticity, deciding how to handle
this pre-factor $\beta$ is non-trivial.

A simple example of entropy production that is difficult to handle in
the BF picture is the production of light from stars, and the
subsequent thermalization of that light by dust. This process converts
high energy photons into many lower energy ones. The energy density is
conserved, yet the temperature decreases. Using $\sigma =
\frac{\rho+P}{T}$, we can see that here the entropy will be greatly
increased, yet the equation of state remains the same.  

For an extreme example of non-adiabaticity, a long period of reheating
can have an effective equation of state equivalent to non-relativistic
matter, and subsequently upon final exit of reheating gains the
stiffer equation of state for radiation.  At constant energy densities,
$\sigma = \beta \rho^{\frac{1}{1+w}}$ increases enormously.
A hypothetical universe dominated by matter with a decay time shorter
than $t_{\Lambda}$ would be another such scenario.  It is not so
obvious in either example exactly when one should calculate the
post-inflation entropy that is to be compared to the entropy bound for
the final state of matter.  Whichever choice one makes, any of the
non-adiabatic scenarios will have a lowered initial entropy compared
to an analogous adiabatic cosmology with similar final matter configuration.  

For an illustration of how these choices reflect truly different versions of the BF
holographic picture, consider the above example of a universe dominated by unstable matter followed by a decay back to radiation.  Inflation would end with 
\begin{align}
S \sim e^{3N_e}H_I^{-3}H_I^{\frac{2}{1+w}} = e^{3N_e}H_I^{-1},
\end{align}
whereas the bound would be\footnote{There is entropy production in the decay of the matter, but it is not clear whether it would be enough to saturate the bound for a new equation of state. In particular, this relation assumes that all the matter decayed to radiation, which is then in thermal equilibrium.}
\begin{align}
S < H_{\Lambda}^{-3 + \frac{2}{1+w}} = H_{\Lambda}^{-3/2}.
\end{align}
This allows 
\begin{align}
3N_e + (- \ln H_I ) &< (-3/2)\ln H_{\Lambda} \\
 N_e &< (3/2)\ln(\frac{H_{\Lambda}}{H_I})^{-1}+ (1/6) \ln{H_I^{-1}},
 \end{align}
a much larger bound than the one finds in the purely 
radiation-dominated universe:
\begin{align}
N_e < (3/2)\ln(\frac{H_{\Lambda}}{H_I})^{-1}.
\end{align}
The result of the extra inflation in this calculation is that some
radiation produced after reheating never reenters the apparent horizon by the
time of Lambda domination, since the subsequent evolution under matter
domination does not increase $H^{-1}$ quickly enough to ``catch" the
biggest scales produced at the start of inflation.   

A universe with matter that never reenters the maximal observer's
horizon is one that is physically different from the one described by
dSE.  One could of course resolve the difference by carefully formulating the
BF bound so as to anticipate the exact degree of non-adiabaticity within the 
cosmological solution chosen.  Or one could interpret the larger bound on e-foldings
in the BF picture as simply a high estimate, less stringent than could be obtained with
more careful analysis.  It is worth noting that as formulated the dSE bound is not sensitive to
these particular concerns over entropy production.  For this reason any attempt to match the BF and dSE
bounds more exactly in realistic cosmologies will generally require clarifications or modifications to the setup of BF.  
There is no guarantee that such modifications will be defensible without the guidance of some organizing principle.
We find the geometric ideas within dSE offer a useful approach. 

There is a further problem with accommodating entropy production within the BF framework. We obtain the bound on the e-foldings of inflation using only the entropy at reheating.
If we are to accept that entropy only present at a later time can influence the evolution of the universe during inflation, then we see no reason that all sources of future entropy  should not similarly affect inflation. These considerations make the BF bound seem rather ad hoc. A geometric picture such as dSE does not suffer from this problem (neither does BF with the requirement of adiabaticity). 

\subsection{Black hole formation}

In our universe, the dominant contribution to the entropy is the
de~Sitter horizon, and after that, black
holes~\cite{Frampton:2008mw}. If included in the BF calculation, the
entropy of black hole formation would completely invalidate the
adiabatic approximation, requiring either a very different approach or
resulting in a substantially weaker 
bound on the total number of e-foldings. However, Banks
and Fischler explicitly exclude black hole entropy, and the universe
is treated as if it were uniform density (effectively replacing the
mass in black holes with a contribution to the uniform cosmological
energy and entropy density). But if we are to truly
ignore black hole entropy as being hidden behind the horizon, then
they would instead contribute no entropy at all.  
Carefully implementing the BF prescription by properly accounting for
the hidden black hole matter will therefore \emph{reduce} the counted
entropy of our universe with every black hole formed. In any case, in
our universe the mass fraction of black holes is small, so this
approximation makes no practical difference in the comparison to the
dSE case.\footnote{We leave open the question of potentially
  substantial differences in universes with a large black hole
  fraction.} But the formal 
exclusion is a real effect on the entropy counting and thus the
calculation of entropy bounds; in comparison, the large-scale
geometric approach of the dSE picture is unaffected by local
replacement of matter with black holes.

\subsection{Tolman-Oppenheimer-Volkoff equation \label{sec:tov}}


We have discussed various methods for estimating the maximum
entropy of a {\em homogeneous} region with a particular fluid equation of
state, but Banks and Fischler ask what bounds arise for the most general
fluid configuration, and use the Tolman-Oppenheimer-Volkoff (TOV)
equation to go beyond the homogeneous case.  The  
TOV equation determines the equilibrium solution
which optimizes the amount of entropy which may be stuffed within a volume
without collapsing to a black hole.\footnote{Although the TOV equation assumes spherical
  symmetry and hydrostatic equilibrium, a series of papers~\cite{Hartle1967, Katz:1975aq,
    Sorkin:1981wd, Fang:2013oka} 
showed that the TOV solutions are also the maximal entropy solutions
for any configuration of a fluid, independent of symmetry (in cases
that do not collapse).}  Here we further extend Banks and 
Fischler's TOV work to the case of a universe with a cosmological
constant. 


The TOV equation 
\begin{align}
\frac{dp}{dr} = -\frac{(\rho + p)(Gm(r) + 4\pi G r^3 p)}{r(r - 2Gm(r))}
\end{align}
has families of solutions with different central densities, one of which is the homogeneous solution~\cite{Axenides:2013hrq}. Including the cosmological constant, the TOV equation
becomes the TOV-$\Lambda$ equation, where $p = p + p_\Lambda$ and
$m(r) = \int_0^r 4\pi {r'}^2 \rho dr' + 4\pi/3 \rho_\Lambda r^3$. 


Our universe is extremely homogeneous on the largest scales, so any
non-uniform TOV solution is a poor fit to our
universe. However, we would like to know if the BF bound actually
favors a different universe.  Applying the dSE and BF
pictures in these universes may also better illuminate differences
between the two approaches. 
%

%

\begin{figure}
\begin{center}
\includegraphics[scale = 0.78]{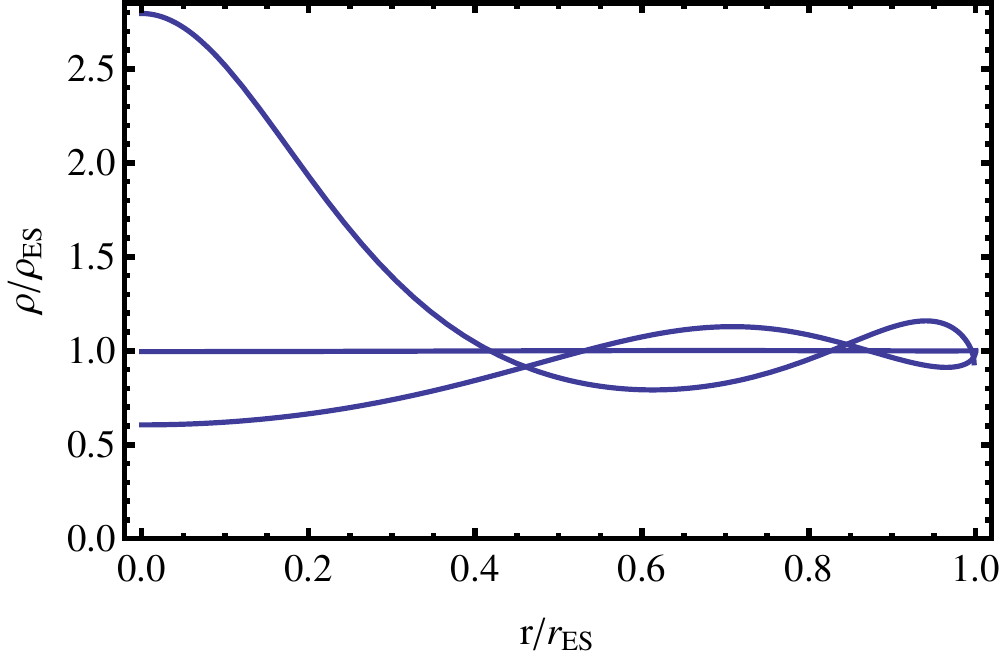}
\caption{The variation of energy density as a function of
  distance coordinate for several TOV-$\Lambda$ universes. Note how the energy 
  density tends to oscillate around the fixed value that would
  correspond to the Einstein Static universe. The behavior of these
  curves is governed by the central density and the equation of state
  parameter $w$ of the fluid. Fluids with $w \to 0$ oscillate more
  than fluids with $w \to 1$. Shown are curves for a fluid with
  $w=0.3$. The straight central line corresponds to the Einstein
  static universe, which is a member of the family of solutions. These
  plots are normalized by the energy density of this Einstein static
  universe ($\rho_{ES}$) and the apparent horizon of the Einstein
  static universe ($r_{ES}$). \label{fig:tovlam}} 
\end{center}
\label{tovlam}
\end{figure}

The solutions to the TOV-$\Lambda$ equation resemble an Einstein
static universe. They are closed static universes of finite size which
extend out to the apparent horizon and  
have the average energy density in matter equal to roughly twice the
energy density of the cosmological constant. 
Figure~\ref{fig:tovlam} shows some illustrative solutions.
%


We find that the inhomogeneity of the TOV solutions does not
significantly increase the entropy over the Einstein static
universe, so in the end there is no point in considering the
inhomogeneous case. Furthermore, the Einstein static universe does
not have an asymptotic de~Sitter future so we cannot directly apply
either the BF or dSE 
bounds. As we will discuss in the next section, we can however examine
``loitering universes'' with slightly less matter content than the
Einstein static universe. 






\subsection{Other cosmologies \label{sec:loiter}}

\begin{figure}
\begin{center}
\includegraphics[scale = 0.78]{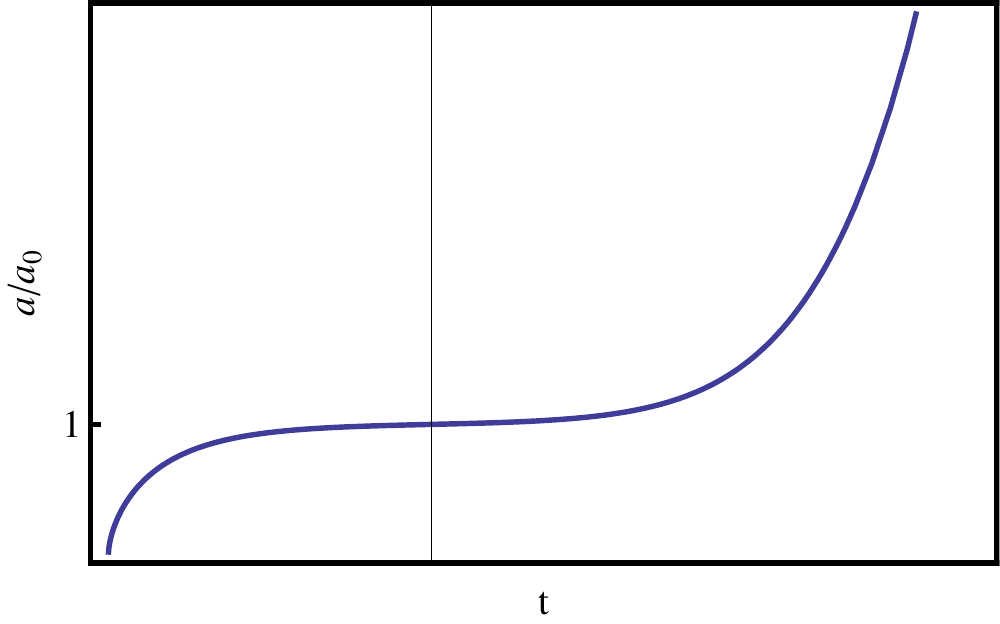}
\caption{The evolution of the scale factor vs. time for a loitering
  universe. Because the expansion slows down during the loitering
  phase, the Hubble length becomes very
  large. Here $a_0$ is set arbitrarily at the inflection point. \label{fig:loiter_a_t}}  
\end{center}
\end{figure}

\begin{figure}
\begin{center}
\includegraphics[scale = 0.78]{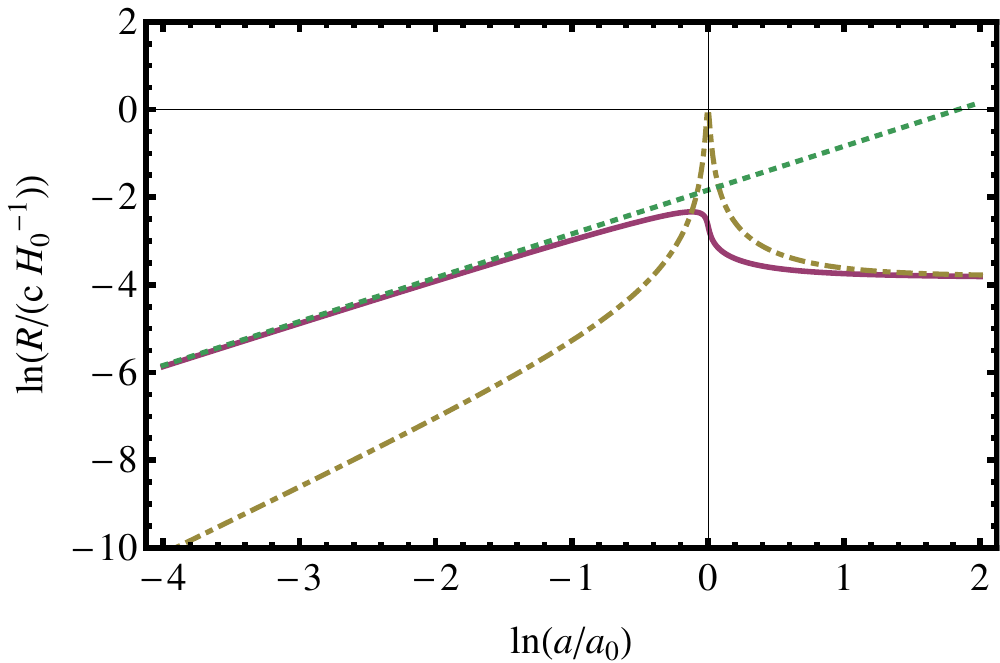}
\caption{Evolution of length scales $R$ for the loitering universe
  shown in Fig.~\ref{fig:loiter_a_t}
  vs. $a$ shown near the era of
  cosmological constant domination.  Plotted are the Hubble length
  (dot-dashed) and 
  the past horizon $h_P$ of an event late in the de~Sitter era (solid). The dotted
  line shows the approximation $h_P \propto a$. Notice that the
  loitering phase causes the horizons to be larger, allowing more of
  the universe to be observed. The upturns in the curves of Hubble
  length and past horizon near $a=a_0$ depend on the duration of the
  loitering phase, and can be increased arbitrarily (though the level
  of fine tuning also increases similarly).  \label{fig:loiter}}   
\end{center}
\end{figure}

\begin{figure}
\begin{center}
\includegraphics[scale = 0.78]{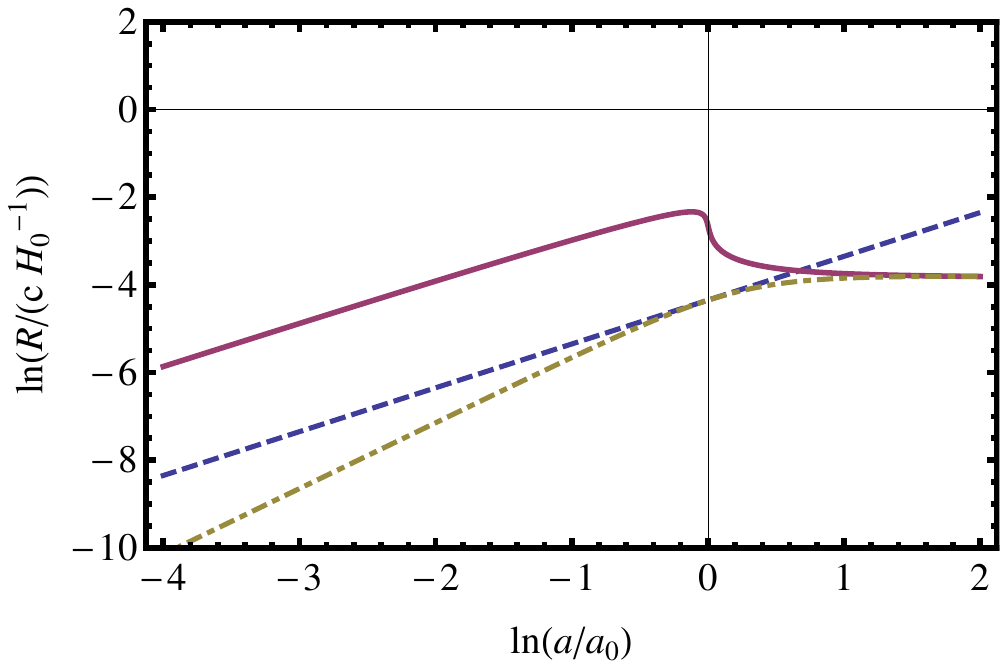}
\caption{More information about the same loitering universe shown in
  Fig.~\ref{fig:loiter}, also plotted vs. scale factor $a$ and shown
  near the era of cosmological constant domination.  Plotted are the apparent
  horizon (dot-dashed) and the past horizon of an event late in the de~Sitter era (solid). The dashed line tracks the comoving volume of space
  with the maximal entropy that can be contained in the fluid cavity
  of BF assuming adiabaticity. Increasing the duration of the
  loitering phase of this cosmology increases the allowed e-foldings
  in dSE arbitrarily, while having little effect on the BF
  bound. This illustrates how the BF and dSE bounds can become
  arbitrarily different in a loitering universe. \label{fig:loiterAH}}   
\end{center}
\end{figure}

In other cosmologies, the discrepancy between the dSE and BF bounds can become more pronounced. In the case of a
loitering universe (Fig.~\ref{fig:loiter_a_t}) the Hubble
constant becomes very small for an extended period of time before heading on
to de~Sitter expansion. During the loitering phase of slow expansion, these are approximately Einstein Static universes and their large curvature causes the Hubble length to deviate substantially from the apparent horizon. The slow expansion causes the Hubble horizon and consequently the past horizon to become very large, as shown in Fig.~\ref{fig:loiter}; however, the apparent horizon is largely unaffected as shown in Fig.~\ref{fig:loiterAH}. 



If we fine tune the ratios of densities of fluid components and the cosmological
constant, an observer can see an arbitrarily large volume of the universe
within her past horizon. This is troublesome for holographic ideas in
general, and especially we see that the dSE picture would have an
arbitrarily weak limit on e-foldings in an arbitrarily fine-tuned
loitering cosmology. Furthermore, the picture of BF would still have a
strong bound on e-foldings, in spite of being able to observe an
arbitrarily large region of space. This would indicate an arbitrarily large observable
entropy\footnote{There would be negligible cosmological redshift within an arbitrarily large observable region because of the slow expansion during the loitering phase.}, in spite of reaching an asymptotic de~Sitter final state. In
our view neither the BF nor the dSE pictures as currently described
in the literature can be applied without further elaboration to the
loitering cosmologies.  
These cosmologies, while they do not describe our universe, are
interesting and useful as test cases because they allow much more
entropy to be observed while still possessing the same final
asymptotic de~Sitter state. And it is important to note that even a
small-curvature universe of the correct sign has a little bit of this
loitering behavior and the more finely tuned solutions are smoothly connected 
to ours in model parameters.  Thus, in a systematic application
of these holographic principles to some ensemble of cosmologies 
the dSE and possibly also BF approaches could favor 
the loitering direction.



\section{Should the bounds be saturated?}
We have directed our analysis to the task of comparing upper bounds on inflation that arise
in the BF and dSE pictures for cosmology.  A separate question is: how close to this bound should we expect a
universe described by one of these pictures to be?  In the dSE picture
there are some reasons to expect a typical universe to be near the
bound; essentially, the mechanism starting inflation makes it
\emph{more} likely to start high on the inflaton potential.  In the
BF picture things are less clear.  Its statement as a bound on entropy
encourages us to think about maximization because of our experience
with the second law of thermodynamics.  But we are also used to thinking of the
universe as having an extremely low entropy initial state.  It
is actually only
because BF exclude 
gravitational and horizon entropy that we can even begin thinking
about the universe as being near a kind of entropy maximum.  Without
additional principles (such as we have in the dSE case) we do not see a
particular reason that the BF bound need be near
saturation. 

A closely related question is: if we interpret observations of our own
universe within the BF or dSE picture,  should we expect the inflation
experienced by our own universe to saturate these bounds?
The added wrinkle is 
that observations have already established an effective floor on the
number of e-foldings.  Thus the typical bound on 
inflation within the BF picture or the dSE picture amounts to ``just a
few e-foldings more than what we have observed''.  This apparent
coincidence is really the coincidence that we are living at a time
that makes us nearly ``maximal observers'' so we can already see most of
what we will ever see.  At least in the BF case, it is
observation-based priors that drive us close to saturating the bound.
More generally both of these models have the feature that the bounds
are not far above the minimum amount of inflation we expect based on
observations. 

\section{Conclusions}
We have shown that the entropy bound of Banks and Fischler and that of
dSE coincide for a very restrictive set of assumptions and a
simplified cosmology. Yet closer investigation reveals that 
even this result requires approximations within the
models, and indeed the conceptual and practical
differences between them are minimized by the
choice of cosmology. Attempting to perform the
comparison on a cosmology more closely resembling
our own (with its multiple equations of state or
failures of adiabaticity) raises many technical issues
that in aggregate call into question how fundamental the 
correspondence is between these two approaches.

Examining even more exotic cosmologies as test
cases merely heightens these issues, and moreover
shows that the project of implementing either
approach as a consistent, rigorous principle across
cosmologies is not quite as straightforward as it
might appear. While the phrasing of the BF bound in terms of
entropy sounds pleasingly universal, the details of its implementation
rely heavily on the cosmological history of the
universe to which it applies. As we have seen
while attempting this implementation, it is roughly
possible to map the BF picture onto the dSE picture by carefully
working backwards to entropy from geometric notions in which 
the dSE picture is originally phrased. 
Because these geometric ideas are more robust under
variations in cosmological history, we ultimately
find them a more practical and compelling basis
for formulating a predictive holographic principle
for finite universes with inflation. Moreover, the
unexpected complexities arising from examining
unusual cosmologies such as the loitering universes
suggest a need to further sharpen such a principle.

\begin{acknowledgments}
We thank T. Banks and W. Fischler for helpful discussions.  This work
was supported in part by DOE Grant DE-FG03-91ER40674.
\end{acknowledgments}

\bibliography{AA}
\bibliographystyle{apsrev}
\end{document}